\documentclass{aastex}
\usepackage{spr-astr-addons}
\usepackage{url}
\usepackage{color}

\RequirePackage{color}

\usepackage{color}

\begin{document}

\title{Spin Down of Rotating Compact Magnetized Strange Stars in General Relativity}

\slugcomment{} \shorttitle{Spin Down of Rotating Strange Stars}
\shortauthors{Ahmedov et al.}

\author{B.~J. Ahmedov\altaffilmark{1,2,3}} \and
\author{B.~B. Ahmedov\altaffilmark{2,4}} \and
\author{A.~A. Abdujabbarov\altaffilmark{1,2,5}}

\altaffiltext{1}{Institute of Nuclear Physics,
        Ulughbek, Tashkent 100214, Uzbekistan}
\altaffiltext{2}{Ulugh Begh Astronomical Institute,
Astronomicheskaya 33, Tashkent 100052, Uzbekistan}
\altaffiltext{3}{International Centre for Theoretical Physics,
Strada Costiera 11, 34151 Trieste, Italy}
\altaffiltext{4}{National University of Uzbekistan, Tashkent
100174, Uzbekistan}
\altaffiltext{5}{ZARM, University Bremen, Am Fallturm, 28359
Bremen, Germany}

%\altaffiltext{2}{IUCAA, Post Bag 4, Pune 411 007, India}

%\begin{flushleft}
%PACSes:{{04.40.-b}{95.30.Qd}{95.30.Sf}}\\
%\end{flushleft}

\begin{abstract}
\noindent We find that in general relativity slow down of the
pulsar rotation due to the magnetodipolar radiation is more faster
for the strange star with comparison  to that for the neutron star
of the same mass. Comparison with astrophysical observations on
pulsars spindown data may provide an evidence for the strange star
existence and, thus, serve as a test for distinguishing it from
the neutron star.
\end{abstract}


\section{Introduction}

The study of electromagnetic fields  of magnetized compact objects
is an important task for several reasons. First  we obtain
information about such stars through their observable
characteristics, which are closely connected with electromagnetic
fields inside and outside the relativistic stars. Magnetic fields
play an important role in the life history of majority
astrophysical objects especially of compact relativistic stars
which possess surface magnetic fields of $10^{12}~\textrm{G}$ and
$\sim 10^{14}~\textrm{G}$ in the exceptional cases for
magnetars~\citep[see e.g.][]{dt92,td93}. The strength of compact
star's magnetic field is one of the main quantities determining
their observability, for example as pulsars through the
magneto-dipolar radiation. Electromagnetic waves radiated from the
star determine energy losses from the star and therefore may be
related with such observable parameters as period of pulsar and
it's time derivative. The second reason is that we may test
various theories of gravitation through the study of compact
objects for which general relativistic effects are especially
strong. Considering different matter for the stellar structure one
may investigate the effect of the different phenomena on evolution
and behavior of stellar interior and exterior magnetic fields.
Then these models can be checked through comparison of theoretical
results with the observational data. The third reason may be seen
in the influence of stellar magnetic and electric field on the
different physical phenomena around the star, such as
gravitational lensing and motion of test particles.

The majority of neutron stars are known to have large angular
velocities, and in the case of radio pulsars one can directly
measure their speed of rotation. It is also observed that, on
average, their rotation tends to slow down with time, a phenomenon
that is explained by emission of electromagnetic waves or, in some
conditions, by the emission of gravitational waves or other
processes. This should be the case during most of the life of the
neutron star when it is observed as pulsar. Since
1967~\citep{hetal68} pulsars play a role of relativistic
astrophysical laboratory where the fast rotation, the extreme
density of matter and the high magnetic field are realized.
{Topical problems in the physics of and basic facts about neutron
stars are reviewed by~\citet{potekhin}.}

Neutron stars provide a natural laboratory to study extremely
dense matter. In the interiors of such stars, the density can
reach up to several times the nuclear saturation density
$n_0\simeq 0.16~\textrm{fm}^{-3}$. At such high densities quarks
could be squeezed out of nucleons to form quark matter. The true
ground state of dense quark matter at high densities and low
temperatures remains an open problem due to the difficulty of
solving nonperturbative quantum chromodynamics (QCD). It has been
suggested that strange quark matter that consists of comparable
numbers of u, d, and s quarks may be the stable ground state of
normal quark matter. This has led to the conjecture that the
family of compact stars may have members consisting entirely of
quark matter (so-called strange stars) and/or members featuring
quark cores surrounded by a hadronic shell (hybrid stars). The
physics of strange matter is reviewed for example by
\cite{g00,w99,m99,w05,wetal09,x03,hpya07,as10,sa10,sa2011}.

The possibility of existence of stable self-bound strange matter
could have important consequences for neutron stars: some compact,
dense stars could be strange stars. Discovery of a strange star in
the universe would be a confirmation of the validity of our
present theory of the structure of matter. In view of this, we
should look for the signatures of strange stars. Univocal and
detectable signature of strange star would be a key to its
possible detection. Low mass strange stars are much smaller than
neutron stars. There is no lower limit for strange star mass.

Observationally, it is very challenging to distinguish the various
types of compact objects, such as the strange stars, hybrid stars,
and ordinary neutron stars. Newly born strange stars are much more
powerful emitters of neutrinos than neutron stars. Their early
cooling behavior is dominated by neutrino emission which is a
useful probe of the internal composition of compact stars. Thus,
cooling simulations provide an effective test of the nature of
compact stars. However, many theoretical uncertainties and the
current amount of data on the surface temperatures of neutron
stars leave sufficient room for speculations.  However, this
property is also characteristic of the neutron star with a large
quark core.

Another useful avenue for testing the internal structure and
composition of compact stars is astroseismology, i.e., the study
of the phenomena related to stellar vibrations. Pulsations of
newly born strange star are damped in a fraction of second: after
this time copious neutrino flux from them should not show
pulsating features. Unfortunately, the same would be true for the
neutrino emission from a neutron star with a large quark core.

Photon cooling of bare strange star could be significantly
different from that of neutron stars. If quark surface is not an
extremely poor emitter of photons, then absence of insulating
crust could lead to a relatively fast photon cooling. In
principle, a well established upper limit of surface temperature
of neutron star-like object of known age, which is well below the
estimates for an object with crust, could be a signature of a bare
strange star. A neutron star-like object with crust which is
$10~\textrm{y}$ old cannot have a surface temperature lower than
$10^6~\textrm{K}$. Unfortunately, surface (black-body) emission
flux decreases as a fourth power of temperature and at the present
day with  X-ray satellite detector it is very difficult to detect
an object of $10~\textrm{km}$ radius at $100~\textrm{pc}$ (typical
distance to the nearest observed point X-ray sources) if its
surface black body temperature is quite low. However the recent
observation of radio-quiet, X-ray bright, central compact objects
(CCOs) shows that the spectra of these objects can very well be
described by a one- or two-component blackbody model, which would
indicate unusually small radii ($\sim 5~\textrm{km}$) for these
objects~\cite{becker09}. Such small radii can only be explained in
terms of self-bound stellar objects like strange quark matter
star.

The search for the detectable signatures of strange stars should
be continued. After all, chances of producing strange matter in
our laboratories are negligibly small compared to those for its
creation during $10^{10}$ years in the immense cosmic laboratory
of the universe.

Here in this short note we will be concerned with the
possibility to distinguish neutron star from the strange star from
the spin down of pulsar. We compare the spindown features of neutron star and strange
star and discuss if astrophysical observations could be useful to prove
the validity of the strange matter hypothesis.

If strange stars are born in some supernova explosions then
because of enormous electric conductivity of strange matter they
should possess huge frozen-in magnetic field~\citep{h87}. In this
respect strange stars could be used as models of pulsars. The
calculation of electric conductivity $\sigma$ of strange matter
has been done by~\cite{hpya07}. Charge transport in strange matter
is dominated by quarks. The value of conductivity $\sigma$ is
determined by the color-screened QCD interaction $\sigma = 10^{29}
\textrm{T}_{10}^{-2}~s^{-1}$ and it is only several times larger
than electric conductivity of normal neutron star matter of the
same density and temperature. In the case of neutron star matter
the charge carriers are electrons.

The magnetic field of strange stars will decay due to ohmic
dissipation of currents strange matter. For the dipole magnetic
field the decay time is \citep[see e.g.,][]{ll}
\begin{equation}
\tau_{D} \cong \frac{4\sigma R^2}{\pi c^2}
    \ ,
\end{equation}
and for $T< 10^9~\textrm{K}$ and $R = 10~\textrm{km}$ one can get
$\tau_{D}> 4 \cdot 10^{11}~\textrm{yr}$. The ohmic dissipation of
magnetic field in a strange star is thus negligible.

Now we plan to study the spin-down of a rotating strange star due
to magnetodipolar electromagnetic emission. Assume that the
oblique rotating magnetized star is observed as radio pulsar
through magnetic dipole radiation. Then the luminosity of the
relativistic star in the case of a purely dipolar radiation, and
the power radiated in the form of dipolar electromagnetic
radiation, is given by \citet{ra04}
\begin{equation}
\label{dipole_energy_loss} L_{em} = \frac{\Omega_{_R}^4 R^6
\widetilde{B}^2_{0}}{6 c^3}\sin^2\chi
    \ ,
\end{equation}
where {tilde denotes the general relativistic value of the
corresponding quantity, subscript $R$ denotes the value of the
corresponding quantity at $r=R$} and $\chi$ is the inclination
angle between magnetic and rotational axes. {In this report we
will use the spacetime of slowly rotating relativistic star which
in a coordinate system $(t, r, \theta, \varphi)$ has the following
form:
\begin{eqnarray}
ds^2&=&-e^{2\Phi(r)}dt^2 + e^{2\Lambda(r)}dr^2- 2\omega(r) r^2
\sin^2\theta dtd\varphi
\nonumber\\
\nonumber\\
&&+r^2d\theta^2+r^2\sin^2\theta d\varphi^2\ ,\label{metric}
\end{eqnarray}
where metric functions $\Phi$ and $\lambda$ are completely known
for the outside of the star and given as:
\begin{eqnarray}
e^{2\Phi}=\left(1-\frac{2M}{r}\right)=e^{-2\Lambda}\ ,
\end{eqnarray}
$\omega=2J/r^3$, $J=I(M,~R)$ is the total angular momentum of the
star with total mass $M$ and moment of inertia $I(M,~R)$.}

{For the interior of the star the metric functions strongly depend
on equation of state and was widely discussed in the
literature~\citep[see, e.g.][]{fp10,getal00,pgz00}. In particular,
the function $\Lambda(r)$ is related to mass function $m(r)$:
\begin{eqnarray}
e^{2\Lambda}=\left(1-\frac{2m}{r}\right)^{-1}\ .
\end{eqnarray}
Function  $\Phi(r) $ can be determined  from evaluating the
following integral \citep[see,][for more details]{fp10}:
\begin{eqnarray}
\Phi(r)=\frac{1}{2}\ln\left(1-\frac{2M}{r}\right) -
\int_{r}^{R}\frac{m(x)+4\pi x^3 p(x)}{x^2(1-2m(x)/x)}dx\ ,
\end{eqnarray}
where $p(x)$ is the pressure profile over the parameter $0 \leq
x\leq R$. }

When compared with the equivalent Newtonian expression for the
rate of electromagnetic energy loss through dipolar radiation
\citep{ll87,p68},
\begin{equation}
\label{dipole_energy_loss_newt} (L_{em})_{\rm Newt.} =
    \frac{\Omega^4 R^6 B_0^2}{6 c^3}\sin^2\chi
    \ ,
\end{equation}
it is easy to realize that the general relativistic corrections
emerging in expression (\ref{dipole_energy_loss}) are due partly
to the magnetic field amplification at the stellar surface
\begin{eqnarray}
\label{mf_amplfctn} \frac{{\widetilde B}_0}{B_0} &=&
\frac{{\widetilde B}_0 R^3}{2 \mu} = f_{_R} \ ,\nonumber\\
f_{_R}&=&-\frac{3R^3}{8M^3}\left[\ln N_{_R}^2+\frac{2M}{R}
    \left(1+\frac{M}{R}\right)\right]\ ,
\end{eqnarray}
and partly to the increase in the effective rotational angular
velocity produced by the gravitational redshift
\begin{equation}
\label{grs} \Omega(r) = \Omega_{_R} \frac{N_{_R}}{N} =
    \Omega_{_R} \sqrt{\left(\frac{R-2M}{r-2M}\right)\frac{r}{R}} \
    ,
\end{equation}
i.e. $\Omega = \Omega_{_R} N_{_R}= \Omega_{_R} \sqrt{1-2M/R}$ when
$r \rightarrow \infty$, where $N=(1-2M/r)^{1/2}$ is the lapse
function.

Expression (\ref{dipole_energy_loss})  could be used to
investigate the rotational evolution of magnetized neutron stars
with predominant dipolar magnetic field anchored in the crust
which converts its rotational energy into electromagnetic
radiation. First detailed  investigation of general relativistic
effects for Schwarzschild stars has been performed
by~\cite{pgz00}, who have paid special attention to the general
relativistic corrections that needed  to be included for a correct
modeling of the thermal evolution but also of the magnetic and
rotational evolution.

Overall, therefore, the presence of a curved spacetime has the
effect of increasing the rate of energy loss through dipolar
electromagnetic radiation for the strange star with comparison  to
that for the neutron star by an amount which can be easily
estimated to be
\begin{equation}
\label{dipole_energy_loss_cf} \frac{(L_{em})_{_{\rm
SS}}}{(L_{em})_{_{\rm NS}}}=
        \left(\frac{f_{_R}}{N^2_{_R}}\right)^2_{_{\rm
SS}}/\left(\frac{f_{_R}}{N^2_{_R}}\right)^2_{_{\rm NS}}\ .
\end{equation}

The expression for the energy loss (\ref{dipole_energy_loss}) can
also be used to determine the spin-evolution of a pulsar that
converts its rotational energy into electromagnetic radiation.
Following the simple arguments proposed more than thirty years ago
\citep{p68,go69,go69b}, it is possible to relate the
electromagnetic energy loss $L_{em}$ directly to the loss of
rotational kinetic energy $E_{\rm rot}$ defined as
\begin{equation}
\label{rot_ent} E_{\rm rot} \equiv \frac{1}{2}
    \int d^3{\bf x} \sqrt{\gamma} e^{-\Phi(r)}
    \rho (\delta v^{\hat \varphi})^2
     \ ,
\end{equation}
where $\rho$ is the stellar energy density and factor $\gamma$ is
defined as follow:
$$
\gamma=\left[-g_{00}\left(1+g_{ik}\frac{\delta v^{i}\delta
v^{k}}{g_{00}}\right)\right]^{-1/2}\simeq e^{-\Phi}\ ,
$$
$\delta v^{i} = dx^{i}/dt$ is the three velocity of conducting
medium defined by~\citet{ra04}, $g_{\alpha\beta}$ is the
components of the spacetime metric~(\ref{metric}), Greek indices
run from 0 to 3, Latin indices from 1 to 3, and hatted quantities
$(\delta v^{\hat i})$ are defined in the orthonormal frame carried
by the static observers in the stellar interior.

\citet{fp10} have shown that having generated an equation of state
it is easy to find pressure $p(r)$ and density $\rho(r)$ profile.
In most physical models considered in the literature, for example
in the paper of~\citet{fp10} the ratio $p/\rho$ {in the crust}
lies in the interval $(4 \div 5)\times 10^{-3}$. As one moves
towards the center of the star this ratio could increase to as
large as $10\%$. However the moment of the inertia is very
sensitive to the matter density/pressure in the crust of the star
and one can assume that the condition $p/\rho\ll 1$ will
approximately satisfy, where $p$ is the pressure of the stellar
matter. This is suitable in the most physical situations and one
can introduce the general relativistic moment of the inertia of
the star as~\citep[see,
e.g.][]{ra04,abdik09,fp10,sn03,worleyetal08}:
\begin{equation}
\label{mom_inrt} {\widetilde I} \equiv \int d^3{\bf x}
\sqrt{\gamma} e^{-\Phi(r)}
    \rho r^2\sin^2\theta\ ,
\end{equation}
whose Newtonian limit gives the well-known expression $I \equiv
({\widetilde I})_{\rm Newt.} = \frac{2}{5} M R^2$, the energy
budget is then readily written as
\begin{equation}
\label{en_budgt} {\dot E}_{\rm rot} \equiv
    \frac{d}{dt}\left(\frac{1}{2} {\widetilde I} \Omega^2\right)
    = - L_{em} \ .
\end{equation}
Of course, in enforcing the balance (\ref{en_budgt}) we are
implicitly assuming all the other losses of energy (e.g. those to
gravitational waves) to be negligible. This can be a reasonable
approximation except during the initial stages of the pulsar's
life, during which the energy losses due to emission of
gravitational radiation will dominate because of the steeper
dependence on the angular velocity (i.e. ${\dot E}_{_{\rm
GW}}\propto \Omega^6$).

%Assuming the spin-down is induced by the magnetic dipole
%radiation, the evolution of the rotation frequency  is given by
%where I is the stellar moment of inertia,  is the magnetic dipole
%moment.

Expression (\ref{en_budgt}) can also be written in a more useful
form in terms of the pulsar's most important observables: the
period $P$ and its time derivative ${\dot P}\equiv dP/dt$. In this
case, in fact, using expression (\ref{dipole_energy_loss}) and
(\ref{en_budgt}), it is not difficult to show that
\begin{equation}
\label{ppdot} (P {\dot P})_{_{\rm SS}} =
\left(\frac{f^2_{_R}}{N^4_{_R}}\right)_{_{\rm
SS}}\left(\frac{f^2_{_R}}{N^4_{_R}}\right)_{_{\rm NS}}^{-1}
\frac{{\widetilde I}_{_{\rm NS}}}{{\widetilde I}_{_{\rm SS}}}
    (P {\dot P})_{_{\rm NS}}
    \ .
\end{equation}

Also in this case it is not difficult to realize that general
relativistic corrections will be introduced through the
amplification of the magnetic field and of the stellar angular
velocity, as well as of the stellar moment of inertia.

\begin{table*}\label{tab1}
%table caption is above the table
\caption{ The dependence of the ratio $(P {\dot P})_{_{\rm SS}}/(P
{\dot P})_{_{\rm NS}}$ from the different parameters of the
compact object: mass (in units of solar mass), radii and moment of
inertia of the Strange ($R_{\rm SS}$, $I_{\rm SS}$) and Neutron
($R_{\rm NS}$, $I_{\rm NS}$) stars. Data for strange and neutron
stars are obtained from the recent paper of \citet{bagchi10}. }
       % Give a unique label
\begin{center}
{\begin{tabular}{@{}ccccc@{}} \hline\noalign{\smallskip}
 ${(P {\dot P})_{_{\rm SS}}}/{(P
{\dot P})_{_{\rm NS}}}$ & 4.34463 & 4.53723 & 5.1094 & 6.16863 \\
\noalign{\smallskip}\hline\noalign{\smallskip}
 ${{M}/{M_\odot}}$ & 1.2 &  1.3 & 1.4 & 1.5  \\
%\hline
 $R_{\rm SS}\ ,\  {\rm km}$ & 7.48 & 7.62 & 7.69 & 7.68 \\
%\hline
 $R_{\rm NS}\ ,\  {\rm km}$ & 11.75 & 11.72 & 11.7 & 11.68 \\
%\hline
 $I_{\rm SS}\ ,\  \times 10^{45}\  {\rm gm \ cm^2}$ & 0.65 & 0.74 & 0.825 & 0.9 \\
%\hline
 $I_{\rm NS}\ ,\  \times 10^{45}\  {\rm gm \ cm^2}$ & 1.08 & 1.2 & 1.36 & 1.72 \\
  \noalign{\smallskip}\hline
\end{tabular}}
\end{center}
\end{table*}

Considering slowly rotating magnetized neutron star one can see
that the general relativistic corrections emerging in expression
(\ref{dipole_energy_loss}) will be partly due to the magnetic
field amplification at the stellar surface and partly to the
increase in the effective rotational angular velocity produced by
the gravitational redshift.

 General-relativistic treatment
for the structure of external and internal stellar magnetic fields
including numerical results has shown that the magnetic field is
amplified by the monopolar part of gravitational field depending
on the compactness of the relativistic star. Thus for a given compact star,
the effects of general relativity on electromagnetic luminosity can be
characterized only by the single compactness parameter $M/R$ which is
different for the neutron and strange star.

Let us mention the so called canonical neutron star model used by
many authors. This artificial model does not imply any specific
EOS, but just assumes the typical values of $M$ and $R$: $M =
1.4~M_\odot$, $R = 10~\textrm{km}$. Using the data for the mass,
the radius, the moment of inertia of neutron stars and strange
stars from the recent paper \citet{bagchi10} we have calculated
the ratio of spin down of neutron star to one of the strange star
on the base formula (\ref{ppdot}) for the compact stars of the
different masses. Results are summarized in the table \ref{tab1}
from where one can see that the strange star is spinning down
approximately 5 times faster that the neutron star.

The pulsar period $P$ versus period derivative ${\dot P}$
is astrophysically measured (see, e.g. \citet{hl06}, \citet{m04})
distinguishes the different classes of pulsars.
According to the astrophysical observations the majority of pulsars have the periods of ~1 s and
period derivatives of $10^{-16}$ to $10^{-14}$. Since period derivatives are in the range of
about two orders one may conclude that the neutron stars have less period derivative with
compare to the strange stars.

In  the present paper we considered the general relativistic
effects on the electromagnetic luminosity of a rotating magnetic strange star
which is  produced due to the rotation of the strange
star with the inclined dipolar magnetic
field configuration. It is shown that the effect
of compactness of strange star on the electromagnetic power loss of the
star is non-negligible (may have the order of tens percents of the
value for the neutron star) and may help in future in
distinguishing the strange star model via pulsar timing
observations.

As an important application of the obtained results we have
calculated energy losses of slowly rotating strange star and found
that the strange star will lose more energy than typical rotating
neutron star in general relativity. The obtained dependence may be
combined with the astrophysical data on pulsar period slowdowns
and be useful in further investigations of the possible
detection/distinguish of the strange stars.

Recently Lavagetto et al. (2005a,2005b) have shown important role
of the general-relativistic effects in the evolution of low-mass
X-ray binaries hosting a neutron star and of millisecond binary
radio pulsars. In particular the formula for the energy released
by magnetodipolar rotator obtained by~\citet{ra04} has been
applied for the angular momentum loss by the neutron star at the
pulsar phase of the evolution. The general relativistic formulas
for the electromagnetic energy released by oscillating star can be
used for the oscillation energy loss by the neutron star in the
binary system when it is observed through quasiperiodic
oscillations (QPOs). Development of model of QPOs in binary system
hosting magnetized oscillating neutron star is other possible
extension of this research.

\section*{Acknowledgments}

This research is supported in part by the UzFFR (projects 1-10 and
11-10) and projects FA-F2-F079, FA-F2-F061 of the UzAS and by the
ICTP through the OEA-PRJ-29 project. We wish to thank Viktoriya
Morozova for many helpful discussions and comments. We thank the
IUCAA for the hospitality where the research has been conducted.
ABJ thanks the TWAS for the associateship grant. AAA thanks the
German Academic Exchange Service (DAAD) for financial support.

%\section*{References}

\bibliographystyle{spr-mp-nameyear-cnd}

\end{document}